\documentclass[preprintnumbers,amsmath,amssymb,floatfix,11pt,prd,superscriptaddress,nofootinbib]{revtex4}
\usepackage{graphicx}
\usepackage{epsfig}
\usepackage{bm}
\usepackage{amsfonts}

\def\ben{\begin{equation}}
\def\een{\end{equation}}

 \def\bd{\begin{document}} \def\ed{\end{document}}
\def\ds{\documentstyle} \let\fr=\frac \let\bl=\bigl \let\br=\bigr
\let\Br=\Bigr \let\Bl=\Bigl
\let\bm=\bibitem
\let\na=\nabla
\let\pa=\partial \let\ov=\overline
\newcommand{\be}{\begin{equation}}
\newcommand{\ee}{\end{equation}}
\def\ba{\begin{array}}
\def\ea{\end{array}}
\def\ft#1#2{{\textstyle{\frac{\scriptstyle #1}{\scriptstyle #2} } }}
\def\fft#1#2{{\frac{#1}{#2}}}
\def\del{\partial}
\def\vp{\varphi}
\def\sst#1{{\scriptscriptstyle #1}}
\def\oneone{\rlap 1\mkern4mu{\rm l}}
\def\td{\tilde}
\def\wtd{\widetilde}
\def\ie{{\it i.e.\ }}
\def\dalemb#1#2{{\vbox{\hrule height .#2pt
        \hbox{\vrule width.#2pt height#1pt \kern#1pt
                \vrule width.#2pt}
        \hrule height.#2pt}}}
\def\square{\mathord{\dalemb{6.8}{7}\hbox{\hskip1pt}}}
\newcommand{\ho}[1]{$\, ^{#1}$}
\newcommand{\hoch}[1]{$\, ^{#1}$}
\newcommand{\bea}{\setlength\arraycolsep{2pt} \begin{eqnarray}}
\newcommand{\eea}{\end{eqnarray}}
\newcommand{\ra}{\rightarrow}
\newcommand{\lra}{\longrightarrow}
\newcommand{\Lra}{\Leftrightarrow}
\newcommand{\bp}{\tilde \beta^\prime}
\newcommand{\tr}{{\rm tr} }
\newcommand{\Tr}{{\rm Tr} }
\def\0{{\sst{(0)}}}
\def\1{{\sst{(1)}}}
\def\2{{\sst{(2)}}}
\def\3{{\sst{(3)}}}
\def\4{{\sst{(4)}}}
\def\5{{\sst{(5)}}}
\def\6{{\sst{(6)}}}
\def\7{{\sst{(7)}}}
\def\8{{\sst{(8)}}}
\def\m{{\sst{(m)}}}
\def\n{{\sst{(n)}}}
\def\cA{{{\cal A}}}
\def\cB{{{\cal B}}}
\def\cF{{{\cal F}}}
\def\cG{{{\cal G}}}
\def\cH{{{\cal H}}}
\def\tV{\widetilde V}
\def\tW{\widetilde W}
\def\tH{\widetilde H}
\def\tE{\widetilde E}
\def\tF{\widetilde F}
\def\tA{\widetilde A}
\def\im{{{\rm i}}}
\def\tY{{{\wtd Y}}}
\def\ep{{\epsilon}}
\def\vep{{\varepsilon}}
\def\bD{{{\bar D}}}
\def\R{{{\mathbb R}}}
\def\C{{{\mathbb C}}}
\def\H{{{\mathbb H}}}
\def\CP{{{\mathbb C}{\mathbb P}}}
\def\RP{{{\mathbb R}{\mathbb P}}}
\def\Z{{{\mathbb Z}}}
\def\bA{{{\mathbb A}}}
\def\bB{{{\mathbb B}}}
\def\bC{{{\mathbb C}}}
\def\bD{{{\mathbb D}}}
\def\bE{{{\mathbb E}}}
\def\bZ{{{\mathbb Z}}}
\def\Re{{{\frak{Re}}}}
\def\Im{{{\frak{Im}}}}
\def\cosec{{\,\hbox{cosec}\,}}
\def\Gm{{\Gamma_{\!\! -}}}
\def\Gp{{\Gamma_{\!\! +}}}
\def\stan{{standard }}
\def\nonstan{{supernumerary }}
\def\p{{\partial}}
\def\kdel#1{{\fft{\del}{\del#1}}}

\def\bog{{Bogomolny }}
\def\om{{\omega}}
\newcommand{\un}{\underline}
\def\R{\hbox{{\rm I}\kern-0.2em{\rm R}\kern0.2em}}%mathematical R for reals
\def\D{\hbox{{\rm I}\kern-0.2em{\rm D}\kern0.2em}}
\def\a{\alpha} \def\o{\omega} \def\w{\wedge}
\def\b{\beta}         \def\rf{\rfloor}
\def\e{{\rm e}}
\def\ld{\lambda} \def\Ld{\Omega}
\def\d{{\rm d}}
\def\dsy{\displaystyle}
\def\de{\delta}
\def\ep{\epsilon}
\def\g{\gamma}
\def\be{\begin{equation}}
\def\ee{\end{equation}}
\def\X{{\cal X}} \def\U{{\cal U}}
\def\p{\partial}
\def\({\left(}
\def\){\right)}
\def\[{\left[}
\def\]{\right]}
\def\bc{\begin{center}}
\def\ec{\end{center}}
\def\EL{Euler-Lagrange}
\def\ph{\phantom}
\def\wh{\widehat}
\def\n{\noindent}
\newcommand{\nnr}{\nonumber \\}
\newcommand{\pd}{\partial}
\newcommand{\ud}{\textrm{d}}
\newcommand{\dTH}{T^{\prime \, 0}_\textrm{H}}
\newcommand{\dOi}{\Omega^{\prime \, 0}_i}
\newcommand{\bx}{{\bf x}}
\begin{document}

\title{Bianchi Type I Cosmology in Generalized Saez-Ballester Theory via Noether Gauge Symmetry }
\author{\textbf{Mubasher Jamil}}
\email{mjamil@camp.nust.edu.pk}
 \affiliation{Center for Advanced Mathematics and Physics (CAMP), National University of Sciences and Technology (NUST), H-12, Islamabad, Pakistan}

\affiliation{Eurasian International Center for Theoretical
Physics,Eurasian National University, Astana 010008, Kazakhstan}

\author{\textbf{Sajid Ali}}
\email{sajid_ali@mail.com} \affiliation{School of Electrical
Engineering and Computer Sciences (SEECS), National University of Sciences and Technology (NUST), H-12, Islamabad, Pakistan}

\author{\textbf{D. Momeni}}
\email{d.momeni@yahoo.com} \affiliation{Eurasian International
Center for Theoretical Physics,Eurasian National University, Astana
010008, Kazakhstan}
\author{\textbf{R. Myrzakulov}}
\email{rmyrzakulov@csufresno.edu}\affiliation{Eurasian International Center for Theoretical Physics,Eurasian National University, Astana 010008,
Kazakhstan}
\begin{abstract}
\vspace*{1.5cm} \centerline{\bf Abstract} \vspace*{6mm}

In this paper, we investigate the 
generalized  Saez-Ballester scalar-tensor theory of gravity via Noether gauge symmetry (NGS) in the background of Bianchi type I cosmological spacetime. We
start with the Lagrangian of our model and calculate its gauge
symmetries and corresponding invariant quantities. We obtain the potential function for the scalar
field in the exponential form. For all the symmetries obtained, we
determine the gauge funñtions corresponding to each gauge symmmetry which include constant and dynamic gauge. We discuss cosmological implications of our model and show that it is compatible with the observational data. \vspace{6mm}

\textbf{Keywords:} Bianchi type I spacetime; Cosmology; Noether
symmetry; Saez-Ballester scalar-tensor theory.

\end{abstract}

\maketitle

\newpage

\section{Introduction}

Several cosmological observations indicate that the observable
universe is undergoing a phase of accelerated expansion \cite{perl}.
There are two major approaches to address the problem of cosmic
acceleration: either introducing a `dark energy' component in the universe
and study its dynamics \cite{sahni} or interpreting it as a failure of general relativity (GR) and consider modifying GR theory, termed as the `modified gravity'
approach \cite{od}. Although both approaches have novel features with
some deep theoretical problems, we here focus only on the modified gravity
approach.

One of the earlier modifications to Einstein's general relativity
was termed Brans-Dicke gravity, in which besides a gravitational
part, a dynamical scalar field was introduced to account for a
variable gravitational constant \cite{dicke}. This modification was
introduced due to lack of compatibility of Einstein's theory with
the Mach's principle. Another cosmologically viable model is the covariant scalar-tensor-vector theory which allows the gravitational constant $G$ \cite{moffat}. The theory can explain successfully galaxy rotation curves and cluster data without non-baryonic dark matter. Moreover the theory is consistent with solar system observational tests. Later Saez \& Ballester \cite{sb} introduced a
scalar-tensor theory of gravity in which metric is coupled to a
scalar field. Here the strength of the coupling between gravity and
the field was governed by a parameter $\omega$. With this
modification, they were able to solve a `missing-mass problem'.
Several aspects of Saez \& Ballester theory in relation to Bianchi
cosmological models have been explored in literature \cite{bb}.

Noether symmetries are the symmetries of the Lagrangian. In
literature, the approach of Noether symmetry is used  to obtain
exact forms of gravitational theories including $f(T)$ gravity,
where $T$ is torsion scalar \cite{f}, $f(R)$ gravity, $R$
being Ricci scalar \cite{j} and scalar-tensor theories \cite{tensor}. Spherically symmetric solutions in $f(R)$ gravity via Noether symmetry were discussed in \cite{caap}. This approach gives a power-law evolutionary form of
scale factor which is consistent with the astrophysical
observations. Moreover the evolution of state-parameter obtained in
such an approach also gives a phantom crossing behavior of dark
energy \cite{c}. The Noether symmetry approach has been applied to
Bianchi cosmological models in literature: Capozziello et al \cite{Capozziello}
investigated the Bianchi universes via Noether symmetries. Camci \&
Kucukakca \cite{camci} studied the Noether symmetries of Bianchi
type I, III and Kantowski-Sachs spacetimes in scalar coupled
theories. They obtained the exact solutions for potential
functions, scalar field and the scale factors, see also \cite{sn}
which is a similar work as \cite{camci}. Scalar-tensor theories have been investigated via Noether symmetry \cite{modak} but via NGS approach, the analysis is not reported earlier. The application of Noether theorem in higher order theory of gravity turned out to be a powerful tool to find the solution of the field equations and physically reasonable solutions like power law inflation have been discussed in \cite{moda}. The NS approach has been applied to pure gravity model with variable cosmological constant $\Lambda$ and gravitational constant $G$ in \cite{alf}.

In this paper, we consider a Bianchi type I spacetime in the
framework of Saez-Ballester theory of gravity. We set up a
Lagrangian in which the metric variables and scalar potential
play the role of dynamical variables. The Lagrangian is so
constructed that its variation with respect to the metric components
and the scalar potential leads to the correct equations of motion.
We explicitly calculate the form of scalar potential by demanding
the Lagrangian admits the desired Noether symmetry. Unlike the usual
approach to obtain Noether symmetries $\mathcal{L}_XL=0$ as followed
in \cite{j}, we employ the full Noether Gauge Symmetry Condition ($X^{[1]}L + (D_{t}\xi)L = D_{t}G(t,\phi, A, B, C)$)
introduced earlier in \cite{jamil}. The advantage of this later
scheme is that it yields extra symmetries then the former one, hence
full depth of the theory is realized in this manner.

The plan of the paper is as follows: In section II, we model our
system by writing the Lagrangian and then deriving the dynamical
equations of motion for Bianchi type I spacetime. In section III, we
consider pure vacuum solution and construct a Lagrangian. Using it,
we solve system of coupled differential equations to obtain Noether
gauge symmetries and corresponding invariant quantities. We discuss some cosmological implications of our model in section IV.
 We conclude
in section V.

\section{ The Model }

The metric of   Bianchi model of type I in coordinates
$x^\mu=(t,x,y,z)$ is \cite{bianchi}
%%%%
\be g_{\mu\nu}=diag(1,-A^2(t),-B^2(t),-C^2(t)) \label{metric} \ee
 %%%
 The exact solutions of Einstein field equations based on 
metric (\ref{metric}) have been investigated in detail in the literature \cite{exact,bianchi}.
The geometrical quantities of (\ref{metric}) are the average scale factor $a=\sqrt[3]{ABC}$; the volume of the spacelike
hypersurface, defined by $V=ABC=a^3$; 
the generalized (or mean) Hubble parameter: $H=\frac{1}{3}\sum_i
H_i$ where $H_i=\partial_t \log(A_i)$, $A_i=\{A,B,C\}$. For the isotropic case, $A=B=C$, the mean Hubble parameter converts
to the Friedmann-Robertson-Walker form $H=\partial_t \log(a)$ where $a$ is the scale
factor. In this paper we are interesting to investigate the
anisotropic models in which the cosmology described by metric
(\ref{metric}) with $A\neq B\neq C$.

We consider the case of a homogeneous but anisotropic
Bianchi type-I model with a scalar field $\phi$ based on a non-standard scalar-tensor theory. The action of this model reads \cite{sb}
%%%%
\be S=\frac{1}{16 \pi
}\int\sqrt{-g}d^4
x\Big(R+\frac{\omega}{2}\phi^k\phi_{,\mu}\phi^{,\mu}-V(\phi)\Big),
 \label{action} \ee
 %%%
where $k$ and $\omega$ are arbitrary dimensionless constants. Choosing $k=0$ reduces our model to the 
minimally coupled massless scalar field coupled to Einstein gravity. Different aspects
of this model have been explored in the literature
\cite{models}.
 Varying (\ref{action}) w.r.t the metric
$g_{\mu\nu}$  lead to a generalized Einstein equation
%%%%
\be G_{\mu\nu}=\omega \phi^k(\phi_{,\mu}
\phi_{,\nu}-\frac{1}{2}g_{\mu\nu}\phi^{,\sigma}\phi_{,\sigma})-V(\phi)g_{\mu\nu}.
 \label{feq} \ee
 %%%
The generalized Klein-Gordon equation for scalar field is
%%%%
\be \frac{1}{\sqrt{-g}}\frac{\partial}{\partial
x^\mu}(\sqrt{-g}g^{\mu\nu}\omega\phi^k\phi_{;\nu})=-V'(\phi)+\frac{\omega
k}{2}\phi^{k-1}\phi_{,\sigma}\phi^{,\sigma}.
 \label{sf} \ee
 %%%
Using metric (\ref{metric}) in the field equations (\ref{feq}) and (\ref{sf}), we obtain
%%%%
\be
\sum_{i,j}^3\Big(\frac{\ddot{A}_{i}}{A_i}+\frac{\ddot{A}_{j}}{A_j}+\frac{\dot{A}_{i}}{A_i}\frac{\dot{A}_{j}}{A_j}\Big)=\frac{1}{2}\omega
\phi^k \dot{\phi}^2,
 \label{a1} \ee
 %%%
%%%%
\be \sum_{i,j}^3\frac{\dot{A}_{i}}{A_i}\frac{\dot{A}_{j}}{A_j}=-\frac{1}{2}\omega
\phi^k \dot{\phi}^2+V(\phi).
 \label{a2} \ee
 %%%
The Klein-Gordon equation is
%%%%
\be  \frac{1}{a^3}\frac{d}{d
t}(a^3\omega\phi^k\dot{\phi})=-V'(\phi)+\frac{\omega
k}{2}\phi^{k-1}\dot{\phi}^2.
 \label{a3} \ee
 %%%

\section{NGS Analysis }

We eliminate the terms $\ddot{A}_{i}$ and
obtain the following Lagrangian which is suitable for calculating
the gauge symmetries:
\begin{equation}\label{L}
L (t,\phi, A, B , C , \dot{\phi},\dot{A}, \dot{B}, \dot {C}) =
\left (\frac{\omega}{2}\phi ^{k}  \dot{\phi}^2 - V(\phi) \right)ABC
- 2 (\dot{A}\dot{B}C + A\dot{B}\dot{C} + \dot{A}B\dot{C}), \quad
\omega \neq 0
\end{equation}
Varying the Lagrangian (\ref{L}) w.r.t. $\phi,$ $A$, $B$ and $C$, we get a
 system of Euler-Lagrange equations (or field equations):
\begin{align}
\ddot{\phi} = -\frac{2\phi ABC V_{\phi} + 2 \omega \phi^{k} \dot{\phi}( k ABC \dot{\phi} +  \phi ( \dot{A}BC +A\dot{B}C + AB\dot{C}))} {2 \omega \phi^{k+1} A B C } ,     \\
\ddot{A} = \frac{4 (A \dot{B} \dot{C} -   \dot{A} B\dot{C} -
\dot{A} \dot{B} C) + 2ABCV -w \phi ^{k} \dot{\phi}^2 ABC  }{8 B C},
\\
\ddot{B} = \frac{4 (\dot{A} B\dot{C} - A \dot{B} \dot{C} -  \dot{A}
\dot{B} C) +2ABCV - w \phi ^{k} \dot{\phi}^2 ABC  }{8 A C},
\\
\ddot{C} = \frac{4 (\dot{A} \dot{B} C -  \dot{A} B\dot{C} -  A\dot{B}
\dot{C} ) + 2ABCV -w \phi ^{k} \dot{\phi}^2 ABC  }{8 A B}.
\label{eleqs}
\end{align}
The Noether symmetry is given by
\begin{equation}
X = \xi \frac{\partial}{\partial t} + \eta_{1}
\frac{\partial}{\partial T} +\eta _2 \frac{\partial}{\partial A}
+ \eta_{3} \frac{\partial}{\partial B}+ \eta_{4}
\frac{\partial}{\partial C}
\end{equation}
where the coefficients $\xi,~ \eta_{i}, ~ (i=1,2,3,4)$ are
determined from the Noether symmetry conditions. The first order
prolongation of the above symmetry to the first-order jet space
comprising of all derivatives is
\begin{equation*}
X^{[1]} = X+ \dot{\eta}_{1} \frac{\partial}{\partial \dot{T}} +
\dot{\eta} _2 \frac{\partial}{\partial \dot{A}} + \dot{\eta}_{3}
\frac{\partial}{\partial \dot{B}}+ \dot{\eta}_{4}
\frac{\partial}{\partial \dot{C}}
\end{equation*}
The Noether gauge symmetry condition is \cite{jamil}
\begin{equation}\label{ngs1}
X^{[1]}L + (D_{t}\xi)L = D_{t}G(t,\phi, A, B, C),
\end{equation}
where $G$ is the gauge function. We emphasize here the difference between the Noether and Noether gauge symmetries: In fact the NS is a very special case of NGS i.e. ignoring the gauge function and first prolongation, we find the restricted (or a special form of) Noether symmetry. The set of Noether symmetries is always a subset of Noether gauge symmetries.

The condition (\ref{ngs1}) yields the following system of linear partial differential equations
\begin{equation}
\xi_{{\phi}} =0,
\nonumber 
\end{equation}
\begin{equation}
\xi_{{A}} =0,
\nonumber 
\end{equation}
\begin{equation}
\xi_{{B}} =0,
\nonumber 
\end{equation}
\begin{equation}
\xi_{{C}} =0,
\nonumber 
\end{equation}
\begin{equation}
C\eta_{{3,A}}+B\eta_{{4,A}}=0,
\nonumber \end{equation}
\begin{equation}
C\eta_{{2,B}}+A\eta_{{4,B}}=0,
\nonumber \end{equation}
\begin{equation}
B\eta_{{2,C}}+A\eta_{{3,C}}=0,
\nonumber \end{equation}
\begin{equation}
G_{{A}}+2B\eta_{{4,t}}+2C\eta_{{3,t}} +VABC\xi_{A}=0 ,
\nonumber \end{equation}
\begin{equation}
G_{{B}}+2C\eta_{{2,t}}+2A\eta_{{4,t}} +VABC\xi_{B}=0,
\nonumber \end{equation}
\begin{equation}
G_{{C}}+2A\eta_{{3,t}}+2B\eta_{{2,t}} +VABC\xi_{C} =0,
\nonumber \end{equation}
\begin{equation}
2C\eta_{{3,\phi}}+2B\eta_{{4,\phi}}-\eta_{{1,A}}w{\phi}^{k}ABC =0,
\nonumber \end{equation}
\begin{equation}
2C\eta_{{2,\phi}} +2A\eta_{{4,\phi}}-\eta_{{1,B}}w{\phi}^{k}ABC =0,
\nonumber \end{equation}
\begin{equation}
2B\eta_{{2,\phi}}+2A\eta_{{3,\phi}}-\eta_{{1,C}}w{\phi}^{k}ABC =0,
\nonumber \end{equation}
\begin{equation}
A\eta_{{4,A}}+B\eta_{{4,B}}+C\xi_{{t}} +C\left(
\eta_{{3,B}}-\xi_{{t}} \right)  +C \left( \eta_{{2,A}}-\xi_{{t}}
\right) +\eta_{{4}}=0 ,
\nonumber \end{equation}
\begin{equation}
B\eta_{{2,B}}+C\eta_{{2,C}}+A\xi_{{t}}+A\left(
\eta_{{3,B}}-\xi_{{t}} \right)  + A\left( \eta_{{4,C}}-\xi_{{t}}
\right) +\eta_{{2}}=0,
\nonumber \end{equation}
\begin{equation}
A\eta_{{3,A}}+C\eta_{{3,C}}+ B\xi_{{t}}+B\left(
\eta_{{2,A}}-\xi_{{t}} \right) +B\left(\eta_{{4,C}} -\xi_{{t}}
\right) +\eta_{{3}}=0,
\nonumber \end{equation}
\begin{equation}
kABC\eta_{{1}} + {\phi}(\eta_{{4}}AB +\eta_{{2}}BC  + \eta_{{3}}AC +
\xi_{{t}}ABC+  2\left( \eta_{{1,\phi}}-\xi_{{t}} \right) wABC) =0,
\nonumber \end{equation}
\begin{equation}
G_{{t}}+\eta_{{1}}ABCV_{{\phi}}+
V(ABC\xi_{{t}}+\eta_{{3}}AC+\eta_{{4}}AB+\eta_{{2}}BC) =0,
\nonumber \end{equation}
\begin{equation}
G_{{\phi}}+ ABC(V\xi_{{\phi}} - \eta_{{1,t}}w{\phi}^{k}) =0,
\label{sys}
\end{equation}
corresponding to the gauge functions $G(t,\phi, A, B , C)$. We numerically solve above system of linear
partial differential equations. The potential function $V(\phi)$ is an arbitrary function
whose form will be specified by the determining equations. We have
the following cases.

\subsection{$V(\phi) = 0$} 

The above system
\eqref{sys} altogether yields nine Noether symmetries comprising of
translation, scalings and other symmetries.
\begin{equation}
X_{1} = \frac{\partial}{\partial {t}}, \quad G = \mbox{const.} 
\nonumber\end{equation}
\begin{equation}
X_{2} = \phi^{-k/2}\frac{\partial}{\partial {\phi}}, \quad G = \mbox{const.}
\nonumber\end{equation}
\begin{equation}
X_{3} = t\frac{\partial}{\partial {t}} + C\frac{\partial}{\partial {C}}, \quad G = \mbox{const.} 
\nonumber\end{equation}
\begin{equation}
X_{4} = A\frac{\partial}{\partial {A}} - C\frac{\partial}{\partial {C}}, \quad G = \mbox{const.} 
\nonumber\end{equation}
\begin{equation}
X_{5} = B\frac{\partial}{\partial {B}} - C\frac{\partial}{\partial {C}}, \quad G = \mbox{const.} 
\nonumber\end{equation}
\begin{equation}
X_{6} = \frac{t^2}{2}\frac{\partial}{\partial {t}} + \frac{tA}{3}\frac{\partial}{\partial {A}}+ \frac{tB}{3}\frac{\partial}{\partial {B}}+ \frac{tC}{3}\frac{\partial}{\partial {C}}, \quad G = -\frac{4}{3}ABC 
\nonumber\end{equation}
\begin{equation}
X_{7} = \phi^{-k/2} \ln \left ({\frac{B}{C}} \right ) \frac{\partial}{\partial {\phi}}- \frac{\omega B \phi^{1+k/2}}{k+2}\frac{\partial}{\partial {B}}+ \frac{\omega C \phi^{1+k/2}}{k+2}\frac{\partial}{\partial {C}}, \quad G = \mbox{const.} 
\nonumber\end{equation}
\begin{equation}
X_{8} = \phi^{-k/2} \ln \left ({\frac{A}{C}} \right ) \frac{\partial}{\partial {\phi}}- \frac{\omega A \phi^{1+k/2}}{k+2}\frac{\partial}{\partial {A}}+ \frac{\omega C \phi^{1+k/2}}{k+2}\frac{\partial}{\partial {C}}, \quad G = \mbox{const.} 
\nonumber\end{equation}
\begin{equation}
X_{9} = A\ln \left ({\frac{B}{C}} \right ) \frac{\partial}{\partial {A}}+B \ln \left ({\frac{C}{A}} \right ) \frac{\partial}{\partial {B}} + C\ln \left ({\frac{A}{B}} \right ) \frac{\partial}{\partial {C}}, \quad G = \mbox{const.} 
\nonumber\end{equation}

The corresponding invariants are as follows:
\begin{equation}
 I_{1} = \frac{\omega}{2}\phi ^{k} \dot{\phi}^2 ABC - 2
(\dot{A}\dot{B}C +
A\dot{B}\dot{C} + \dot{A}B\dot{C}), 
\nonumber\end{equation}
\begin{equation}
I_{2} = \phi^{k/2} \omega \dot{\phi}ABC,
\nonumber\end{equation}
\begin{equation}
I_{3} =  2t  \left (A\dot{B}\dot{C}+\dot{A}B\dot{C}  + \dot{A}\dot{B}C \right )- 2C(A\dot{B}+\dot{A}B)  -\frac{1}{2}t \dot{\phi}^2 \omega \phi^{k}ABC ,
\nonumber\end{equation}
\begin{equation}
I_{4} = B(\dot{A}C-A\dot{C}),
\nonumber\end{equation}
\begin{equation}
I_{5} = A(B\dot{C}-\dot{B}C),
\nonumber\end{equation}
\begin{equation}
 I_{6} = t^2\left (A\dot{B}\dot{C}+\dot{A}B\dot{C}  +
\dot{A}\dot{B}C \right ) + \frac{4}{3}ABC  - \frac{4t}{3}(\dot{A}BC
+ A\dot{B}C + A B\dot{C} )
- \frac{t^2}{4}\dot{\phi}^2 \omega \phi^{k}ABC,  
\nonumber\end{equation}
\begin{equation}
I_{7} = \frac{2\omega A (B\dot{C}-\dot{B}C)\phi^{\frac{k+2}{2}}}{k+2} + \omega\phi^{k/2} \dot{\phi}ABC \ln \left ({\frac{B}{C}} \right ),
\nonumber\end{equation}
\begin{equation}
I_{8} = \frac{2\omega B (A\dot{C}-\dot{A}C)\phi^{\frac{k+2}{2}}}{k+2} + \omega\phi^{k/2} \dot{\phi}ABC \ln \left ({\frac{A}{C}} \right ),
\nonumber\end{equation}
\begin{equation} 
 I_{9} = 2(B\dot{C}-\dot{B}C)A\ln{A}-2(A\dot{C}-\dot{A}C)B\ln{B} +
2 (A\dot{B}-\dot{A}B)C\ln{C}.
\nonumber\end{equation}
The first invariant is the Hamiltonian of the system.

\subsection{ $V(\phi) = \frac{2}{3} \alpha^2
\neq 0 $} 

 In this case the system \eqref{sys} again yields
nine Noether symmetries of which six NSs $X_{1}, X_{4}, X_{5},
X_{7}, X_{8},X_{9} $ are the same as above. The additional two NSs
are
\begin{equation}
\tilde{X}_{3} = e^{\alpha t} \frac{\partial}{\partial {t}} + \frac{\alpha e^{\alpha t}}{3} \left (  A\frac{\partial}{\partial {A}}+  B\frac{\partial}{\partial {B}}+ C\frac{\partial}{\partial {C}} \right ), \quad G = -\frac{4}{3}ABC \alpha ^2 e^{\alpha t}\nonumber\end{equation}
\begin{equation}
\tilde{X}_{6} = e^{-\alpha t} \frac{\partial}{\partial {t}} - \frac{\alpha e^{-\alpha t}}{3} \left (  A\frac{\partial}{\partial {A}}+  B\frac{\partial}{\partial {B}}+ C\frac{\partial}{\partial {C}} \right ), \quad G = -\frac{4}{3}ABC \alpha ^2 e^{-\alpha t}. 
\nonumber\end{equation}
The corresponding invariants are as follows:
\begin{equation}
\tilde{I}_{3} =  {{\rm e}^{\alpha t}} \left(  4{\alpha}^{2}ABC + 12 (\dot{A}\dot{B}C + A\dot{B}\dot{C} + \dot{A}B\dot{C}) - 8 \alpha (A\dot{B}C + AB\dot{C} + B \dot{A}C) - 3 \omega {\phi}^{k}{\dot{\phi}}^{2}ABC  \right), \nonumber\end{equation}
\begin{equation}
\tilde{I}_{6} =  {{\rm e}^{-\alpha t}} \left(  4{\alpha}^{2}ABC +
12 (\dot{A}\dot{B}C + A\dot{B}\dot{C} + \dot{A}B\dot{C}) + 8 \alpha
(A\dot{B}C + AB\dot{C} + B \dot{A}C) - 3 \omega
{\phi}^{k}{\dot{\phi}}^{2}ABC  \right).
\nonumber\end{equation}

\subsection{$V(\phi) = \alpha  \exp ({\beta
\phi^{\frac{k+2}{2}}} )$, $k\neq -2$} 

 The system
\eqref{sys} yields five Noether symmetries of which four $X_{1},
X_{4}, X_{5},X_{9} $ are the same as above. An additional NSs is
\begin{align*}
& \tilde{X}_{2} =  t\frac{\partial}{\partial {t}}-
\frac{4\phi^{-k/2}}{\beta (k+2)} \frac{\partial}{\partial {\phi}} +
C\frac{\partial}{\partial {C}}, \quad G = \mbox{const.}
\end{align*}
The corresponding invariant is
\begin{align*}
& \tilde{I}_{2} = 2t \left ( A\dot{B} \dot{C} + \dot{A} B
\dot{C}+\dot{A}\dot{B}C \right ) - 2C\left( A\dot{B} + \dot{A}B
\right) - ABC\left (\alpha t{{\rm e}^{\beta {\phi}^{\frac{k+2}{2}}}}
+ \frac{4\omega \dot{\phi} {\phi}^{k/2}}{\beta (k+2)}  +\frac{\omega
t {\phi}^{k}{\dot{\phi}}^{2}}{2} \right )
\end{align*}
In this approach, we have obtained the exponential potential of the
scalar field. Such exponential potentials are most favored in
cosmology to study dark energy dynamics and fullfill many issues of
dark energy approach, both from a theoretical point of view and in
comparison with available observational data \cite{d}. Therefore in
the present model, we have obtained a solution representing
accelerated expansion and is of immense cosmological interest. The
case with $k=0$ is interesting as it leads to $(V(\phi) = \alpha
\exp ({\beta \phi} ))$. Such potential forms have been used a lot in
phenomenological models of dark energy such as quintessence, phantom
and quintom  \cite{exp}. For this potential, we have acceleration
solution. Further in this case,  when the kinetic term has much
larger pressure than the potential term, then the potential
domination epoch is an attractor solution as long as the potential
is flat, i.e the case $\beta=0$.

\subsection{ $ \mbox{Arbitrary }V(\phi)$} 

 The system
\eqref{sys} gives four Noether symmetries which are the same as
$X_{1}, X_{4}, X_{5}, X_{9}$ in subsection A. Therefore we don't get any
new NGS and corresponding invariant in this case.

%%%%%%%%%%%%%%%%%%%%%
\section{Cosmological implications }
%%%%%%%%%%%%%%%%%%%%
In this section we investigate the general cosmic evolution of the model proposed in section II with an exponential potential given by
\be
V(\phi) = \alpha  \exp ({\beta
\phi^{\frac{k+2}{2}}} ).
\ee
The EoS parameter $w$ and the deceleration parameter $q$ can be constructed analytically. Since $w$ and $q$ depend on metric functions and scalar field, their evolutionary behavior is obtained by numerically solving the Euler-Lagrange  equations (9)-(12) for an appropriate set of the parameters and the initial conditions.

\begin{figure}
\centering
 \includegraphics[scale=0.5] {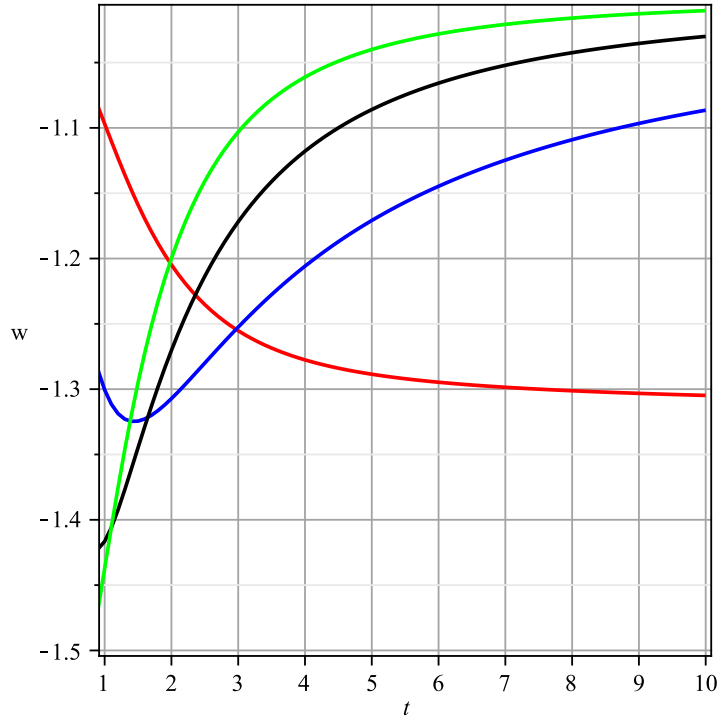}% scale goes from 0 to 1.
  \caption{ Cosmological evolution of $w$
vs time $t$. The model parameters chosen as $\alpha=1,$ $\beta=-1$, $\omega=1$. Curves in various colors correspond to (red, $k=1$), (blue, $k=2$), (black, $k=3$), (green, $k=4$).}
  \label{1.eps}
\end{figure}

\begin{figure}
\centering
\includegraphics[scale=0.5]{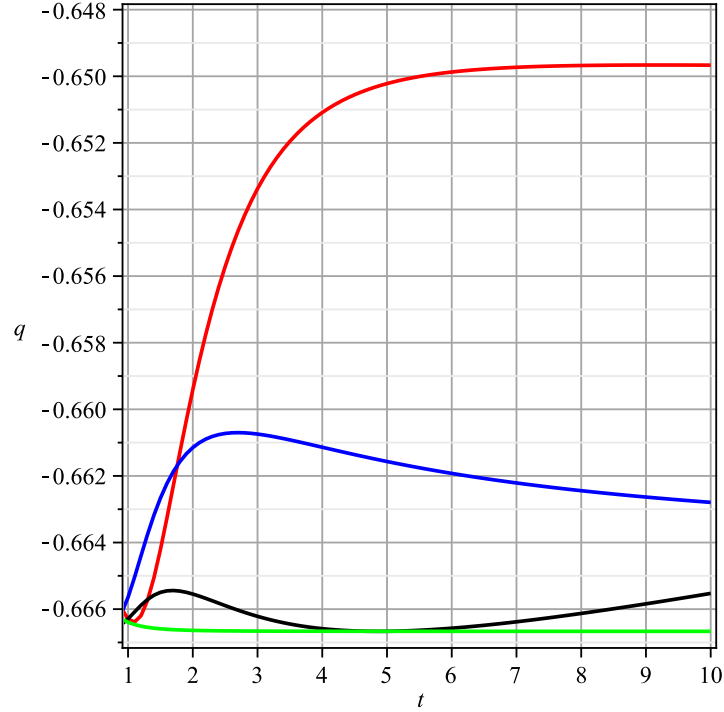}% scale goes from 0 to 1.
  \caption{ Variation of $q$
vs time $t$. The model parameters chosen as $\alpha=1,$ $\beta=-1$, $\omega=1$. Curves in various colors correspond to (red, $k=1$), (blue, $k=2$), (black, $k=3$), (green, $k=4$).}
  \label{2.eps}
\end{figure}

The EoS parameter is constructed using the expressions of total energy density and averaged pressure 
\be
\rho=\frac{1}{2}\omega \phi^k \dot{\phi}^2+V(\phi),
\ee
\be
p=\frac{1}{2}\omega \phi^k \dot{\phi}^2-V(\phi).
\ee
When $k=0$, equations (17), (18) transform to the canonical scalar field model with a rescaling of the field.
The EoS parameter is defined as
\be
w\equiv\frac{p}{\rho}.
\ee
The numerical simulations of $w$ is drawn in figure-1 which shows that $w$ behaves like the phantom form of dark energy. This conclusion is exciting since there exists convincing astrophysical evidence that the observable universe is currently in the phantom phase \cite{caldwell}. In figures, we chose the initial conditions $A(0)=0.2$, $B(0)=0.1$, $C(0)=0.3$, $\phi(0)=1$, $\dot A(0)=1$, $\dot B(0)=1$, $\dot C(0)=1$, $\dot\phi(0)=0.3$.

 Further we calculate the deceleration parameter $q$ using the average scale factor $a$ 
 \be q\equiv-\frac{\ddot{a}a}{\dot{a}^2}=-1-3\Big[\frac{\frac{\ddot{A}}{A}+\frac{\ddot{B}}{B}+\frac{\ddot{C}}{C}-(\frac{\dot{A}}{A})^2-(\frac{\dot{B}}{B})^2-(\frac{\dot{C}}{C})^2}{(\frac{\dot{A}}{A}+\frac{\dot{B}}{B}+\frac{\dot{C}}{C})^2}\Big].
 \ee
Again using the numerical calculation displayed in figure-2, we show that $q$ remains always negative indicating the accelerated expansion of the universe. Note that our model predicts the present value $q_0=-0.67$ which is in good agreement with the astrophysical data \cite{good}.

\section{Conclusion}

In this paper, we investigated the Noether gauge symmetries of a
simple extension of an old model proposed by Saez-Ballester in a
homogenous but anisotropic Bianchi type I backgrounds. We solved the
gauge equations and classified the models depending on potential
function. One of the models is a generalization of the
exponential families which have been used frequently in
phenomenological models of dark energy such as quintessence, phantom
and quintom. By performing numerical simulation of cosmological parameters $w$ and $q$, we demonstrated that the universe lies in the phantom energy dominated phase while the present value of deceleration parameter is compatible with the observations.

\subsection*{Acknowledgment}
M. Jamil and D. Momeni would like to thank the warm hospitality of Eurasian National University, Astana, Kazakhstan where this work was completed. All authors would thank the anonymous referee for the enlightening comments on our paper.

\end{document}